\documentclass[%
 reprint,
superscriptaddress,
 amsmath,amssymb,
 aps,
]{revtex4-1}

\usepackage{graphicx}
\usepackage{epstopdf}

\usepackage{dcolumn}
\usepackage{bm}
\usepackage{hyperref}

\begin{document}

\preprint{APS/123-QED}

\title{Improving the accuracy of atom interferometers with ultracold sources}

\author{R. Karcher}
\author{A. Imanaliev}
\author{S. Merlet}
\author{F. Pereira Dos Santos}

\affiliation{LNE-SYRTE, Observatoire de Paris, Universit\'e PSL, CNRS, Sorbonne Universit\'e, \\61 avenue de l'Observatoire 75014 Paris}

\date{\today}

\begin{abstract}
We report on the implementation of ultracold atoms as a source in a state of the art atom gravimeter. We perform gravity measurements with 10~nm/s$^2$ statistical uncertainties in a so-far unexplored temperature range for such a high accuracy sensor, down to 50~nK. This allows for an improved characterization of the most limiting systematic effect, related to wavefront aberrations of light beam splitters. A thorough model of the impact of this effect onto the measurement is developed and a method is proposed to correct for this bias based on the extrapolation of the measurements down to zero temperature. Finally, an uncertainty of 13 nm/s$^2$ is obtained in the evaluation of this systematic effect, which can be improved further by performing measurements at even lower temperatures. Our results clearly demonstrate the benefit brought by ultracold atoms to the metrological study of free falling atom interferometers. By tackling their main limitation, our method allows reaching record-breaking accuracies for inertial sensors based on atom interferometry.

\end{abstract}

\maketitle

Atom gravimeters constitute today the most mature application of cold atom inertial sensors based on atom interferometry. They reach performances better than their classical counterparts, the free fall corner cube gravimeters, both in terms of short term sensitivity \cite{Hu2013,Gillot2014} and long term stability \cite{Freier2016}. They offer the possibility to perform high repetition rate continuous measurements over extended periods of time \cite{LouchetChauvet2011,Freier2016}, which represents an operation mode inaccessible to other absolute gravimeters. These features have motivated the development of commercial cold atom gravimeters \cite{Menoret2018}, addressing in particular applications in the fields of geophysics. Nevertheless, the accuracy of these sensors is today slightly worse. Best accuracies in the $30-40~\text{nm/s}^2$ range have been reported \cite{LouchetChauvet2011,Freier2016} and validated through the participation of these instruments to international comparisons of absolute gravimeters since 2009 \cite{Jiang2012,Francis2013}, to be compared with the accuracy of the best commercial state of the art corner cube gravimeters, of order of $20~\text{nm/s}^2$ \cite{Microg}. 

\begin{figure}[!ht]
\includegraphics[width=0.45\textwidth]{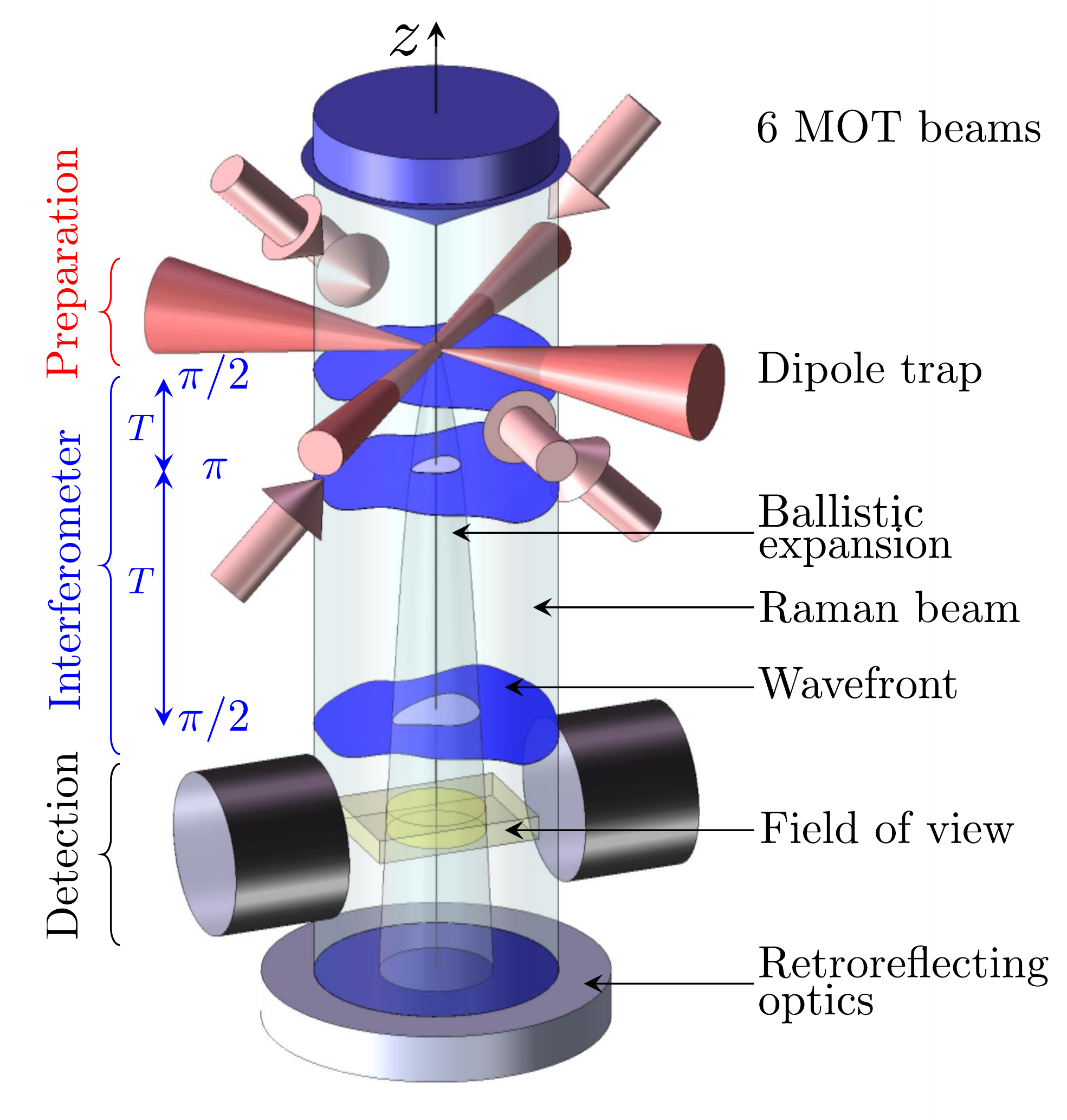}
\caption{(color online) Scheme of the experimental setup, illustrating the effect of wavefront aberrations. Due to their ballistic expansion across the Raman beam, the atoms sample different parasitic phase shifts at the three $\pi/2-\pi-\pi/2$ pulses due to distortions in the wavefront (displayed in blue as a distorted surface). This leads to a bias, resulting from the average of the effect over all atomic trajectories, filtered by finite size effects, such as related to the waist and clear aperture of the Raman beam and to the finite detection field of view.}
\label{fig:scheme}
\end{figure}

The dominant limit in the accuracy of cold atom gravimeters is due to the wavefront distortions of the lasers beamsplitters. This effect is related to the ballistic expansion of the atomic source through its motion in the beamsplitter laser beams, as illustrated in figure \ref{fig:scheme}, and cancels out at zero atomic temperature. In practice, it has been tuned by increasing the atomic temperature \cite{LouchetChauvet2011} and/or by using truncation methods, such as varying the size of the detection area \cite{Schkolnik2015} or of the Raman laser beam \cite{Zhou2016}. Comparing these measurements with measured or modelled wavefronts allows to gain insight on the amplitude of the effect, and estimate the uncertainty on its evaluation. It can be reduced by improving the optical quality of the optical elements of the interferometer lasers, or by operating the interferometer in a cavity \cite{Hamilton2015}, which filters the spatial mode of the lasers, and/or by compensating the wavefront distortions, using for instance a deformable mirror \cite{Trimeche2017}. 

The strategy we pursue here consists in reducing the atomic temperature below the few $\mu$K limit imposed by cooling in optical molasses in order to study the temperature dependence of the wavefront aberration bias over a wider range, and down to the lowest possible temperature. For that, we use ultracold atoms produced by evaporative cooling as the atomic source in our interferometer. Such sources, eventually Bose-Einstein condensed, show high brightness and reduced spatial and velocity spread. These features allow for a drastic increase in the interaction time, on the ground \cite{Dickerson2013} or in space \cite{Altschul2015} and for the efficient implementation of large momentum transfer beam splitters \cite{Szigeti2012,Chiow2011,McDonald2013}. The potential gain in sensitivity has been largely demonstrated (for instance, by up to two orders of magnitude in \cite{Dickerson2013}). But it is only recently that a gain was demonstrated in the measurement sensitivity of an actual inertial quantity \cite{Asenbaum2017}, when compared to best sensors based on the more traditional approach exploiting two photon Raman transitions and laser cooled atoms. Here, implementing such a source in a state of the art absolute gravimeter, we demonstrate that ultracold atom sources also improve the accuracy of atom interferometers, by providing an ideal tool for the precise study of their most limiting systematic effect.
 
We briefly detail here the main features of our cold atom gravimeter. A more detailed description can be found in \cite{LouchetChauvet2011}. It is based on an atom interferometer \cite{Borde2001} based on two-photon Raman transitions, performed on free-falling $^{87}$Rb atoms. A measurement sequence is as follows. We start by collecting a sample of cold or ultracold atoms, which is then released in free fall. After state preparation, a sequence of three Raman pulses drives a Mach Zehnder type interferometer. These pulses, separated by free evolution times of duration $T=80$ ms, respectively split, redirect and recombine the matter waves, creating a two-wave interferometer. The total duration of the interferometer is thus $2T=160$ ms. The populations in the two interferometer output ports $N_1$ and $N_2$ are finally measured by a state selective fluorescence detection method, and the transition probability $P$ is calculated out of these populations ($P=N_1/(N_1+N_2)$). This transition probability depends on the phase difference accumulated by the matter waves along the two arms of the interferometer that is, in our geometry, given by $\Phi= \vec{k}.\vec{g} T^2$, where $\vec{k}$ is the effective wave vector of the Raman transition and $\vec{g}$ the gravity acceleration. Gravity measurements are then repeated in a cyclic manner. Using laser cooled atoms, repetition rates of about 3 Hz are achieved which allows for a fast averaging of the interferometer phase noise dominated by parasitic vibrations. We have demonstrated a best short term sensitivity of $56~\text{nm.s}^{-2}$ at 1~s measurement time \cite{Gillot2014}, which averages down to below $1~\text{nm.s}^{-2}$. These performances are comparable to the ones of the two other best atom gravimeters developed so far \cite{Freier2016,Hu2013}. The use of ultracold atoms reduces the cycling rate due to the increased duration of the preparation of the source. Indeed, we first load the magneto-optical trap for 1~s (instead of 80~ms only when using laser cooled atoms) before transferring the atoms in a far detuned dipole trap realized using a 30~W fibre laser at 1550~nm. It is first focused onto the atoms with a $170~\mu$m waist (radius at 1/e$^2$), before being sent back at a 90$^{\circ}$ angle and tightly focused with a $27~\mu$m waist, forming a crossed dipole trap in the horizontal plane. The cooling and repumping lasers are then switched off, and we end up with about $3\times10^8$ atoms trapped at a temperature of $26~\mu$K. Evaporative cooling is then implemented by decreasing the laser powers from 14.5~W and 8~W to 2.9~W and 100~mW typically in the two arms over a duration of 3~s. We finally end up with atomic samples in the low 100 nK range containing $10^4$ atoms. Changing the powers at the end of the evaporation sequence allows to vary the temperature over a large temperature range, from 50~nK to $7~\mu$K. The total preparation time is then 4.22~s, and the cycle time 4.49~s, which reduces the repetition rate down to 0.22~Hz.
Furthermore, at the lowest temperatures, the number of atoms is reduced down to the level where detection noise becomes comparable to vibration noise. The short term sensitivity is thus significantly degraded and varies in our experiment in the 1200-3000 nm.s$^{-2}$ range at 1~s, depending on the final temperature of the sample.

\begin{figure}[!ht]
\includegraphics[width=0.45\textwidth]{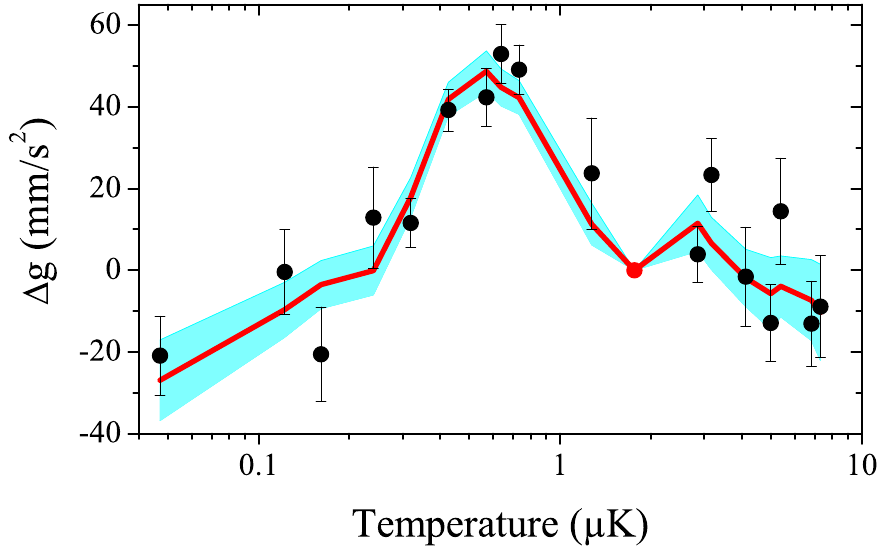}
\caption{(color online) Gravity measurements as a function of the atom temperature. The measurements, displayed as black circles, are performed in a differential way, with respect to a reference temperature of $1.8~\mu$K (displayed as a red circle). The red line is a fit to the data with a subset of five Zernike polynomials and the filled area the corresponding 68\% confidence area.}
\label{fig:data}
\end{figure}

We performed differential measurements of the gravity value as a function of the temperature of the source, which we varied over more than two orders of magnitude. The results are displayed as black circles in figure \ref{fig:data}, which reveals a non-trivial behaviour, with a fairly flat behaviour in the 2-7 $\mu$K range, consistent with previous measurements obtained with optical molasses \cite{LouchetChauvet2011}, and a rapid variation of the measurements below $2~\mu$K. This shows that a linear extrapolation to zero temperature based on high temperature data taken with laser cooled atoms would lead to a significant error. These measurements have been performed for two opposite orientations of the experiment (with respect to the direction of the Earth rotation vector) showing the same behaviour, indicating that these effects are not related to Coriolis acceleration \cite{LouchetChauvet2011}. Moreover, the measurements are performed by averaging measurements with two different orientations of the Raman wavevector, which suppresses the effect of many systematic effects, such as differential light shifts of the Raman lasers that could vary with the temperature \cite{LouchetChauvet2011}.

To interpret these data, we have developed a Monte Carlo model of the experiment, which averages the contributions to the interferometer signal of atoms randomly drawn in their initial position and velocity distributions. It takes into account the selection and interferometer processes, by including the finite size and finite coupling of the Raman lasers, and the detection process, whose finite field of view cuts the contribution of the hottest atoms to the measured atomic populations \cite{Farah2014}. This model is used to calculate the effect of wavefront aberrations onto the gravity measurement as a function of the experimental parameters. For that, we calculate for each randomly drawn atom its trajectory and positions at the three pulses in the Raman beams, and take into account the phase shifts which are imparted to the atomic wavepackets at the Raman pulses: $\delta\phi=k\delta z_i$, where $\delta z_i$ is the flatness defect at the i-th pulse. We sum the contributions of a packet of $10^4$ atoms to the measured atomic populations to evaluate a mean transition probability. The mean phase shift is finally determined from consecutive such determinations of mean transition probabilities using a numerical integrator onto the central fringe of the interferometer, analogous to the measurement protocol used in the experiment \cite{LouchetChauvet2011}. With $10^4$ such packets, we evaluate the interferometer phase shifts with relative uncertainties smaller than $10^{-3}$. We decompose the aberrations $\delta z$ onto the basis of Zernike polynomials $Z_n^m$, taking as a reference radius the finite size of the Raman beam (set by a 28 mm diameter aperture in the optical system). Assuming that the atoms are initially centred on the Raman mirror and in the detection zone, the effect of polynomials with no rotation symmetry ($m \neq 0$) averages to zero, due to the symmetry of the position and velocity distributions \cite{Trimeche2017}. We thus consider here only Zernike polynomials with no angular dependence that correspond to the curvature of the wavefront (or defocus) and to higher order spherical aberrations. 

\begin{figure}[!ht]
\includegraphics[width=0.45\textwidth]{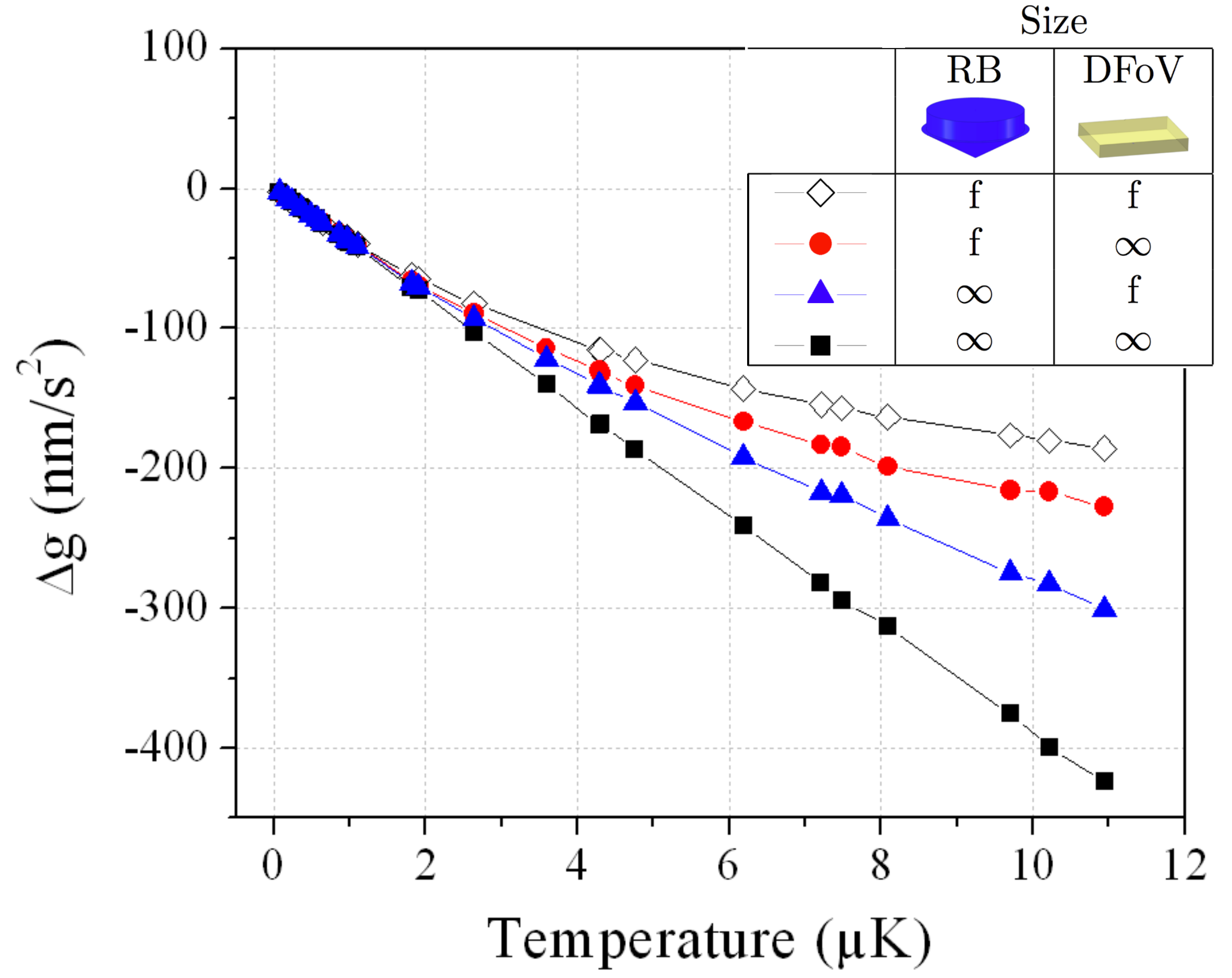}
\caption{(color online) Calculation of the impact of the size of the Raman beam waist (RB) and of the detection field of view (DFoV) on the gravity shift induced by a defocus as a function of the atomic temperature. The peak-to-peak amplitude of the defocus is 20 nm. The results correspond to four different cases, depending on whether the sizes of the Raman beam waist and detection field of view are taken as finite or infinite.}
\label{fig:influence}
\end{figure}

To illustrate the impact of finite size effects, we display in figure \ref{fig:influence} calculated gravity shifts corresponding to different cases, for a defocus ($Z_2^0$) with a peak-to-peak amplitude of $2a_0=20$ nm across the size of the reference radius, which corresponds to $\delta z (r)= a_0(1-2r^2)$, with $r$ the normalized radial distance. The black squares corresponds to the ideal case of infinite Raman laser radius size and detection field of view and give a linear dependence versus temperature. The circles (resp. triangles) correspond to the case of finite beam waist and infinite detection field of view (resp. infinite beam waist and finite detection field of view), and finally diamonds include both finite size effects. Deviations from the linear behaviour arise from the reduction or suppression of the contribution of the hottest atoms. The effect of the finite Raman beam waist is found to be more important than the effect of the finite detection area. Finally, we calculate for this simple study case a bias of -63 nm/s$^2$ at the temperature of $1.8~\mu$K, for a peak-to-peak amplitude of 20 nm. This implies that, at the temperature of laser cooled samples and for a pure curvature, a peak-to-peak amplitude of less than 3 nm ($\lambda$/260 PV) over a reference diameter of 28 mm is required for the bias to be smaller than 10 nm/s$^2$.

\begin{figure}[!ht]
\includegraphics[width=0.45\textwidth]{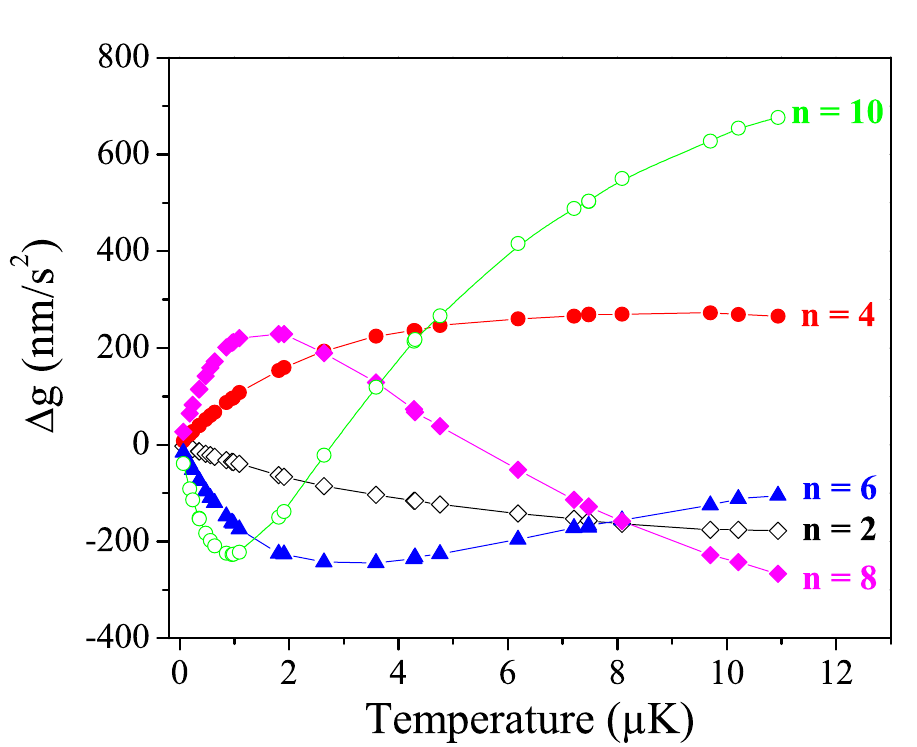}
\caption{(color online) Gravity shifts calculated for the parameters of our experiment, for the five first Zernike polynomials with rotation symmetry $Z_n^0$ (n=2,4,6,8,10) for peak-to-peak amplitudes of 20 nm.}
\label{fig:zernike}
\end{figure}

We then calculate the effect of the 7 first $Z_n^0$ polynomials (for even $n$ ranging from 2 and 14) for the same peak-to-peak amplitude of $2a_0=20$ nm as a function of the atomic temperature. Figure \ref{fig:zernike} displays the results obtained, restricted for clarity to the first five polynomials. All orders exhibit as common features a linear behaviour at low temperatures and a trend for saturation at high temperatures. Interestingly, we find non monotonic behaviours in the temperature range we explore and the presence of local extrema.

Using the phase shifts calculated at the temperatures of the measurements, the data of figure \ref{fig:data} can now be adjusted, using a weighted least square adjustment, by a combination of the contribution of the first Zernike polynomials, which then constitute a finite basis for the decomposition of the wavefront. The adjustment was realized for increasing numbers of polynomials, so as to assess the impact of the truncation of the basis. We give in table \ref{tab:fits} the values of the correlation coefficient $R$ and the extrapolated value at zero temperature as a function of the number of polynomials. We obtain stable values for both $R$ and the extrapolated value to zero temperature, of about $-55~\text{nm/s}^2$ for numbers of polynomials larger than 5. This indicates that the first 5 polynomials are enough to faithfully reconstruct a model wavefront that well reproduces the data. When increasing the number of polynomials, we indeed find that the reconstructed wavefront is dominated by the lowest polynomial orders. The results of the adjustment with 5 polynomials is displayed as a red line in figure \ref{fig:data} and the 68\% confidence bounds as a filled area. The flatness of the reconstructed wavefront at the centre of the Raman laser beam is found to be as small as 20 nm PV (Peak Valley) over a diameter of 20 mm. 
 
\begin{table}
\begin{tabular}{|c|c|c|}
\hline
Number of  & $R$ & Extrapolated value\\
polynomials &  & to zero temperature (nm/s$^2$) \\
\hline
2 & ~0.587~ &  37(17) \\
3 & 0.764 &  09(18) \\
4 & 0.871 & -19(18) \\
5 & 0.963 & {\bf -56(13)} \\
6 & 0.964 & -54(13) \\
7 & 0.964 & -55(15) \\
\hline
\end{tabular}
\caption{Results of the linear adjustments for increasing number of Zernike polynomials. The error bars correspond to $1\sigma$ uncertainties (68\% confidence).}
\label{tab:fits}
\end{table}

The bias due to the optical aberrations at the reference temperature of 1.8 $\mu$K, which corresponds to the temperature of the laser cooled atom source, is thus 56(13) nm/s$^2$. Its uncertainty is three times better than its previous evaluation \cite{LouchetChauvet2011}, which in principle will improve accordingly our accuracy budget.

On the other hand, interatomic interactions in ultracold sources can induce significant phase shifts \cite{Sortais2000,Harber2002} and phase diffusion \cite{Altin2011}, leading to bias and loss of contrast for the interferometer. Nevertheless, the rapid decrease of the atomic density when interrogating the atoms in free fall reduces drastically the impact of interactions \cite{Debs2011,Jannin2015,Abend2016}. To investigate this, we have performed a differential measurement for two different atom numbers at the temperature of 650 nK. The number of atoms was varied from 25000 to 5000 by changing the efficiency of a microwave pulse in the preparation phase, which leaves the spatial distribution and temperature unchanged. We measured an unresolved difference of -7(12) nm/s$^2$. This allows us to put an upper bound on the effect of interactions, which we find lower than 1 nm/s$^2$ per thousand atoms.  

The uncertainty in the evaluation of the bias related to optical aberrations can be improved further by performing measurements at even lower temperatures, which will require in our set-up to improve the efficiency of the evaporative cooling stage. A larger number of atoms would allow to limit the degradation of the short term sensitivity and to perform measurements with shorter averaging times. More, absorption imaging with a vertical probe beam would allow for spatially resolved phase measurements across the cloud \cite{Dickerson2013}, which would allow for improving the reconstruction of the wavefront. The temperature can also be drastically reduced, down to the low nK range, using delta kick collimation techniques \cite{Muntinga2013,Kovachy2015}. In addition to a reduced ballistic expansion, the use of ultracold atoms also offers a better control of the initial position and mean velocity of the source with respect to laser cooled sources, which suffer from fluctuations induced by polarisation and intensity variations of the cooling laser beams. Such an improved control reduces the fluctuations of systematic effects related to the transverse motion of the atoms, such as the Coriolis acceleration and the bias due to aberrations, and thus will improve the long term stability \cite{Abend2016}.

With the above-mentioned improvements, and after a careful re-examination of the accuracy budget \cite{Francis2013}, accuracies better than 10 nm/s$^2$ are within reach. This will make quantum sensors based on atom interferometry the best standards in gravimetry. Furthermore, the improved control of systematics and the resulting gain in stability will open new perspectives for applications, in particular in the field of geophysics \cite{VanCamp2017}. Finally, the method proposed here can be applied to any atomic sensor based on light beamsplitters, which are inevitably affected by distortions of the lasers wavefronts. The improved control of systematics it provides will have significant impact in high precision measurements with atom interferometry, with important applications to geodesy~\cite{Carraz2014,Douch2018}, fundamental physics tests~\cite{Bouchendira2011,Altschul2015,Rosi2014} and to the development of highest grade inertial sensors~\cite{Dutta2016}.

\begin{acknowledgments}
We acknowledge the contributions from X. Joffrin, J. Velardo and C. Guerlin in earlier stages of this project. We thank R. Geiger and A. Landragin for useful discussions and careful reading of the manuscript.
\end{acknowledgments}

\end{document}